\documentclass[aps,prl,reprint,groupedaddress]{revtex4-1}
\usepackage{graphicx} 
\usepackage{dcolumn} 
\usepackage{bm}
\usepackage{hyperref}

\begin{document}

\title{Production of fragments and hyperfragments in antiproton-nucleus collisions}
\author{Zhao-Qing Feng$^{1,2}$}
\email{fengzhq@impcas.ac.cn}

\affiliation{$^{1}$Institute of Modern Physics, Chinese Academy of Sciences, Lanzhou 730000, People's Republic of China            \\
                        $^{2}$Kavli Institute for Theoretical Physics, Chinese Academy of Sciences, Beijing 100190, People's Republic of China}

\date{\today}

\begin{abstract}
Formation mechanism of fragments with strangeness in collisions of antiprotons on nuclei has been investigated within the Lanzhou quantum molecular dynamics (LQMD) transport model. Production of strange particles in the antiproton induced nuclear reactions is modeled within the LQMD model, in which all possible reaction channels such as elastic scattering, annihilation, charge exchange and inelastic scattering in antibaryon-baryon, baryon-baryon and meson-baryon collisions have been included. A coalescence approach is developed for constructing hyperfragments in phase space. The hyperfragments are formed within the narrower rapidities. It has advantage to produce heavier hyperfragments and hypernuclides with strangeness s=-2 (double-$\Lambda$ fragments) and s=1 ($\overline{\Lambda}$-fragments) in antiproton induced reactions.

\begin{description}
\item[PACS number(s)]
21.80.+a, 25.43.+t, 24.10.-i
\end{description}
\end{abstract}

\maketitle

Studies of hypernuclei attract much attention over the past several decades. The interested topics related to hypernuclei are the hyperon-nucleon and hyperon-hyperon interactions, opening a new horizon with strangeness (three-dimensional nuclear chart) in nuclear physics and probing the in-medium properties of hadrons and the inner structure of a nucleus \cite{Gi95, Ha06}. Since the first observation of $\Lambda$-hypernuclides in nuclear multifragmentation reactions induced by cosmic rays in 1950s \cite{Da53}, a remarkable progress has been obtained in producing hypernuclides via different reaction mechanism, such as hadron (pion, kaon, proton) induced reactions, bombarding a fixed target with high-energy photons or electrons, and fragmentation reactions with high energy heavy-ion collisions. It should be mentioned that the delayed fission from the decay of hypernuclei in antiproton annihilations on heavy nuclei were observed for the first time in experiments \cite{Bo86}. The scenario was investigated by the intranuclear cascade (INC) model \cite{Cu90}.

A more localized energy deposition enables the secondary collisions available for producing hyperons. Hyperons produced in antiproton induced reactions can be captured in the potential of nucleon fragments to form hypernuclei. The dynamics of antiproton-nucleus collisions is complicated, which is associated with the mean-field potentials of hadrons in nuclear medium, and also with a number of reaction channels, i.e., the annihilation channels, charge-exchange reaction, elastic and inelastic collisions. The larger yields of strange particles in antiproton induced reactions are favorable to form hypernuclei in comparison to proton-nucleus and heavy-ion collisions. To understand the nuclear dynamics induced by antiprotons, several approaches have been proposed, such as the intranuclear cascade (INC) model \cite{Cu89}, kinetic approach \cite{Ko87}, Giessen Boltzmann-Uehling-Uhlenbeck (GiBUU) transport model \cite{La09}, Statistical Multifragmentation Model (SMM) \cite{Bo95} and the Lanzhou quantum molecular dynamics (LQMD) approach \cite{Fe14}. A number of experimental data were nicely explained within these approaches. Self-consistent description of dynamical evolutions and collisions of antiproton on nucleus is still very necessary, in particular for understanding the fragmentation reactions in collisions of antiprotons on nuclei to form hypernuclei. The production of hypernuclei is associated with the reaction channels of hyperons, and also related to the hyperon-nucleon (HN) and hyperon-hyperon (HH) potentials. The investigation of hypernucleus properties is an essential way for extracting the in-medium information of hyperons.

In the LQMD transport model, the dynamics of resonances ($\Delta$(1232), N*(1440), N*(1535) etc), hyperons ($\Lambda$, $\Sigma$, $\Xi$) and mesons ($\pi$, $K$, $\eta$, $\overline{K}$, $\rho$, $\omega$, $\phi$  etc) is described via hadron-hadron collisions, annihilation reactions of antibaryon-baryon collisions, decays of resonances and transportation in mean-field potentials \cite{Fe11,Fe13}. The temporal evolutions of baryons (nucleons, resonances and hyperons), anti-baryons and mesons in the nuclear collisions are governed by Hamilton's equations of motion. The Hamiltonian of nucleons and nucleonic resonances is derived from the Skyrme energy-density functional and a momentum-dependent potential distinguishing isospin effects has been implemented in the model.

Dynamics of hyperons, anti-baryons and mesons is described within the framework of relativistic mean-field models or chiral perturbation theories. The Hamiltonian is composed of the Coulomb interaction between charged particles and the energy in nuclear medium. The in-medium dispersion relation for hyperons reads as
\begin{equation}
\omega_{Y}(\textbf{p}_{i},\rho_{i})=\sqrt{(m_{Y}+\Sigma_{S}^{Y})^{2}+\textbf{p}_{i}^{2}} + \Sigma_{V}^{Y},
\end{equation}
The hyperon self-energies are evaluated on the basis of the light-quark counting rules, i.e., $\Lambda$ and $\Sigma$ being assumed to be two thirds of nucleon self-energies, the $\Xi$ self-energy being one third of nucleon's ones. Namely, for hyperons $\Sigma_{S}^{\Lambda,\Sigma}= 2 \Sigma_{S}^{N}/3$, $\Sigma_{V}^{\Lambda,\Sigma}= 2 \Sigma_{V}^{N}/3$, $\Sigma_{S}^{\Xi}= \Sigma_{S}^{N}/3$ and $\Sigma_{V}^{\Xi}= \Sigma_{V}^{N}/3$. The antibaryon energy is computed from the G-parity transformation of baryon potential as
\begin{equation}
\omega_{\overline{B}}(\textbf{p}_{i},\rho_{i})=\sqrt{(m_{\overline{B}}+\Sigma_{S}^{\overline{B}})^{2}+\textbf{p}_{i}^{2}} + \Sigma_{V}^{\overline{B}}
\end{equation}
with $\Sigma_{S}^{\overline{B}}=\Sigma_{S}^{B}$ and $\Sigma_{V}^{\overline{B}}=-\Sigma_{V}^{B}$.
The nuclear scalar $\Sigma_{S}^{N}$ and vector $\Sigma_{V}^{N}$ self-energies are computed from the well-known relativistic mean-field model with the NL3 parameter ($g_{\sigma N}^{2}$=80.8, $g_{\omega N}^{2}$=155 and $g_{\rho N}^{2}$=20). The optical potential of baryon or antibaryon is derived from the in-medium energy as $V_{opt}(\textbf{p},\rho)=\omega(\textbf{p},\rho)-\sqrt{\textbf{p}^{2}+m^{2}}$. The relativistic self-energies are used for the construction of hyperon and antibaryon potentials only. A very deep antiproton-nucleus potential is obtained with the G-parity approach being $V_{opt}(\textbf{p}=0,\rho=\rho_{0}) = -655 $ MeV. From fitting the antiproton-nucleus scattering \cite{La09} and the real part of phenomenological antinucleon-nucleon optical potential \cite{Co82}, a factor $\xi$ is introduced in order to moderately evaluate the optical potential as $\Sigma_{S}^{\overline{N}}=\xi\Sigma_{S}^{N}$ and $\Sigma_{V}^{\overline{N}}=-\xi\Sigma_{V}^{N}$ with $\xi$=0.25, which leads to the strength of $V_{\overline{N}}=-164$ MeV at the normal nuclear density $\rho_{0}$=0.16 fm$^{-3}$. It should be noted that the scaling approach violates the fundamental G-symmetry. The antihyperon potentials exhibit strongly attractive interaction in nuclear medium, e.g., the strengths at saturation density being -436 MeV and -218 MeV for $\overline{\Lambda}$ and $\overline{\Xi}$, respectively. The optical potentials will affect the dynamics of hyperons, consequently the production of hypernuclei.

Based on hadron-hadron collisions, to describe the antiproton-nucleus collisions we have further included the annihilation channels, charge-exchange reaction (CEX), elastic (EL) and inelastic scattering as follows \cite{Fe14}:
\begin{eqnarray}
&& \overline{B}B \rightarrow \texttt{annihilation}(\pi,\eta,\rho,\omega,K,\overline{K},\eta\prime,K^{\ast},\overline{K}^{\ast},\phi),
\nonumber \\
&&  \overline{B}B \rightarrow \overline{B}B (\texttt{CEX, EL}),   \overline{N}N \leftrightarrow \overline{N}\Delta(\overline{\Delta}N), \overline{B}B \rightarrow \overline{Y}Y.
\end{eqnarray}
Here the B strands for nucleon and $\Delta$(1232), Y($\Lambda$, $\Sigma$, $\Xi$), K(K$^{0}$, K$^{+}$) and $\overline{K}$($\overline{K^{0}}$, K$^{-}$). The overline of B (Y) means its antiparticle. The cross sections of these channels are based on the parametrization or extrapolation from available experimental data \cite{Bu12}. The annihilation dynamics in antibaryon-baryon collisions is described by a statistical model with SU(3) symmetry inclusion of all pseudoscalar and vector mesons \cite{Go92}, which considers various combinations of possible mesons with the final state from two to six particles \cite{La12}.

\begin{figure*}
\includegraphics[width=16 cm]{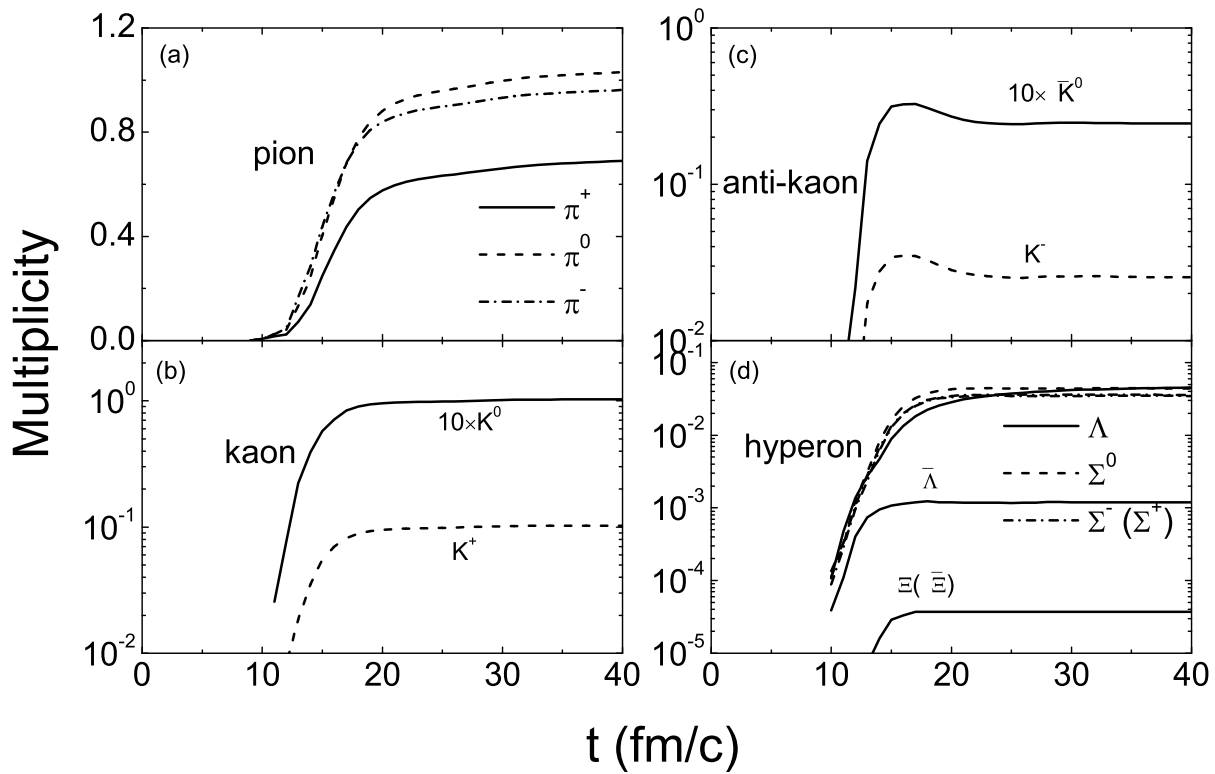}
\caption{ Temporal evolution of particle production in collisions of $\overline{p}$ on $^{40}$Ca at incident momentum of 4 GeV/c.}
\end{figure*}

The emission mechanism of particles produced in antiproton induced reactions is significant in understanding the contributions of different reaction channels associated with antiprotons on nucleons and secondary collisions. Shown in Fig. 1 the temporal evolutions of pions, kaons, antikaons, hyperons and antihyperons in the reaction of $\overline{p}$+$^{40}$Ca at an incident momentum of 4 GeV/c. It is shown that the kaons are emitted immediately after the annihilation in collisions of antibaryons and baryons. The secondary collisions of pions and antikaons on nucleons retard the emissions and even a bump appears for antikaons in the reaction dynamics, i.e., $\pi N\rightarrow \Delta$, $\overline{K}N\rightarrow \pi Y$ etc, which lead to the production of hyperons. At the considered momentum above its threshold energy, e.g., the reaction $\overline{N}N\rightarrow \overline{\Lambda}\Lambda$ (p$_{threshold}$=1.439 GeV/c), production of hyperons are attributed from the direct reaction (annihilation and creation of quark pairs, $u\overline{u} (d\overline{d})\rightarrow s\overline{s}$) and also from the secondary collisions after annihilations in antibaryon-baryon collisions, i.e., meson induced reactions $\pi (\eta, \rho, \omega) N\rightarrow KY$ and strangeness exchange reaction $\overline{K}N\rightarrow \pi Y$.

\begin{figure*}
\includegraphics[width=16 cm]{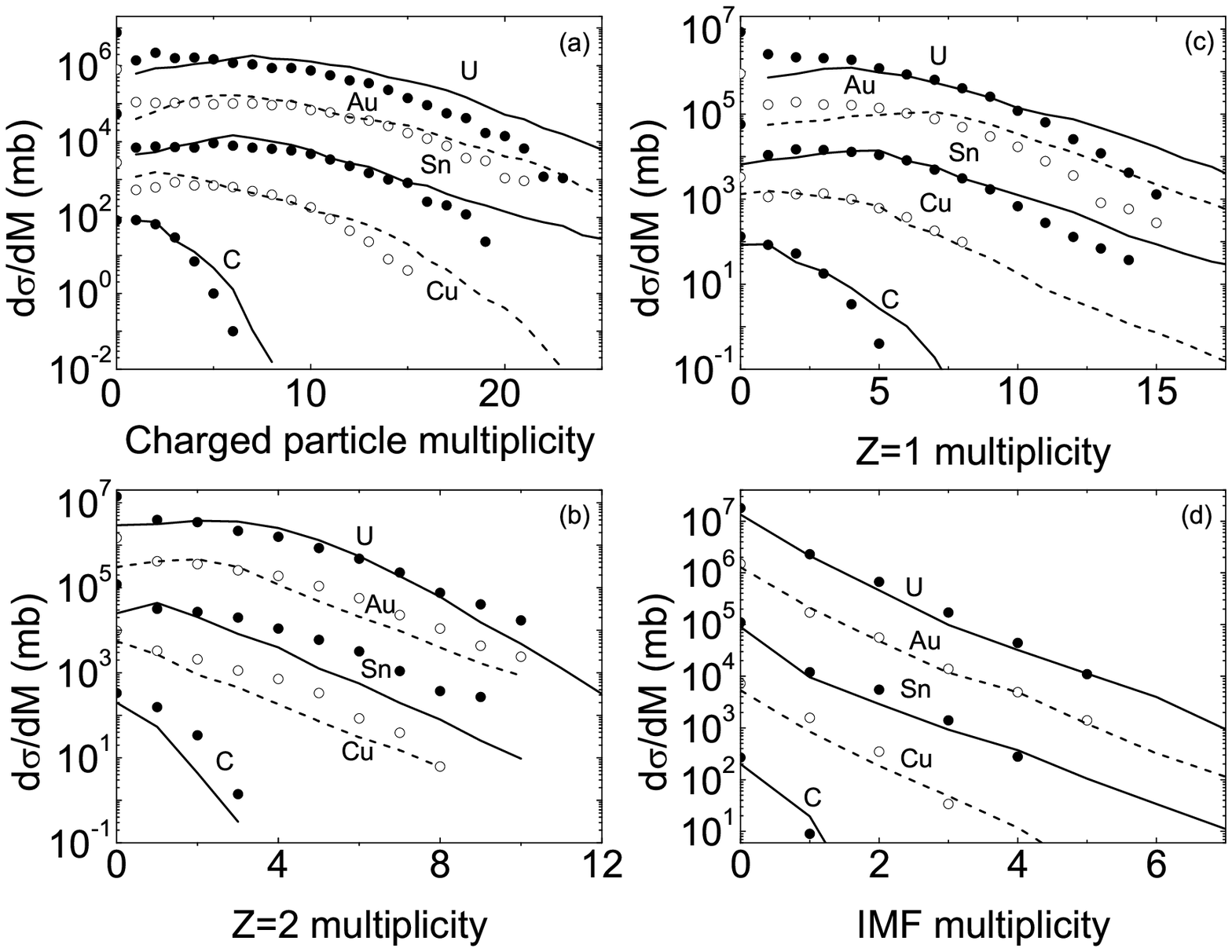}
\caption{ Multiplicity distributions in $\overline{p}$ induced fragmentation reactions on different target nuclides at an incident kinetic energy of 1.22 GeV. The available data are from Ref. \cite{Lo01}. The data and curves are multiplied by succussive powers of 10 for clarity starting from C.}
\end{figure*}

A more localized energy deposition is formed after absorption of antiprotons in nuclei. Roughly, the energy released by stopped antiprotons on Au is similar to that irradiated by 1 GeV protons \cite{Lu02}. Besides a number of mesons emitted after the annihilation of antiprotons in nuclei, target nuclei are excited with the energy deposition, which leads to evaporate nucleons and clusters from the transient nuclei and even to fragmentation reactions. The mass yield distribution could be used to estimate the energy released by antiprotons in nuclei. The fragmentation reactions induced by antiprotons are investigated as shown in Fig. 2. We concentrated on the multiplicity distributions of all charged particles, light charged particles and intermediate mass fragments (IMFs) (3$\leq Z \leq$20). The available data from the low-energy antiproton ring (LEAR) at CERN \cite{Lo01} are compared with calculations from the LQMD transport model combined with the GEMINI statistical decay code for excited fragments \cite{Ch88}. The nuclear dynamics induced by antiprotons is described by the LQMD model. The primary fragments are constructed in phase space with a coalescence model, in which nucleons at freeze-out are considered to belong to one cluster with the relative momentum smaller than $P_{0}$ and with the relative distance smaller than $R_{0}$ (here $P_{0}$ = 200 MeV/c and $R_{0}$ = 3 fm). At the freeze-out, the primary fragments are highly excited. The de-excitation of the fragments is assumed to be isolated without rotation (zero angular momentum). The excitation energy is evaluated as the difference of binding energies between the excited fragment and its groundstate ones. The fragmentation mechanism can be described nicely well with the combined approach.

\begin{figure*}
\includegraphics[width=16 cm]{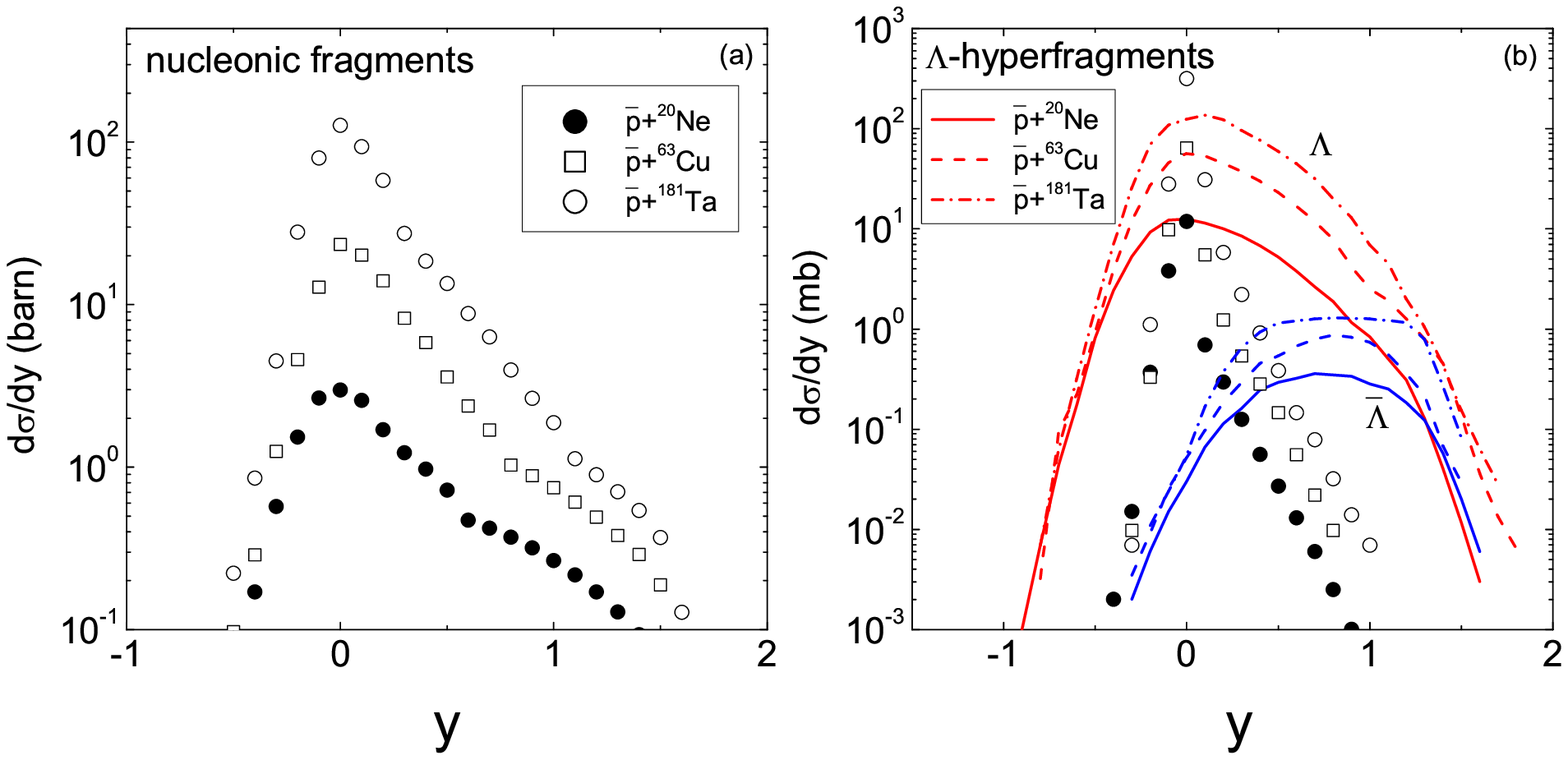}
\caption{(Color online) Rapidity distributions of all fragments and $\Lambda$-hyperfragments formed in collisions of $\overline{p}$ on $^{20}$Ne, $^{63}$Cu and $^{181}$Ta at incident momentum of 4 GeV/c. Hyperons $\Lambda$ and $\overline{\Lambda}$ are indicated for a comparison.}
\end{figure*}

The phase-space structure of nucleonic fragments and hyperfragments produced in antiproton induced reactions would be helpful for the detector management in experiments. Furthermore, the estimation of cross sections for hypernucleus production with an optimal projectile-target combinations and incident energy is favorable to produce hypernuclei with less costs. Shown in Fig. 3 is the rapidity distributions of nucleonic fragments and $\Lambda$-hyperfragments formed in collisions of $\overline{p}$ on $^{20}$Ne, $^{63}$Cu and $^{181}$Ta at an incident momentum of 4 GeV/c. Hyperons $\Lambda$ and $\overline{\Lambda}$ produced in the $\overline{p}$ induced reactions are indicated for a comparison. The hyperons are captured by nucleonic fragments and a narrow rapidity distribution of $\Lambda$-fragments is formed. Here, we assume a larger relative distance ($R_{0}$ = 5 fm) and the relative momentum similar to nucleonic ones ($P_{0}$ = 200 MeV/c) between hyperon and nucleon in constructing a hypernucleus, which is caused from the fact that the weakly bound of hypernucleus with a bigger rms (root-mean-square) radius, e.g., 5 fm rms for $^{3}_{\Lambda}$H and 1.74 fm for $^{3}$He \cite{Ar04}.

\begin{figure*}
\includegraphics[width=16 cm]{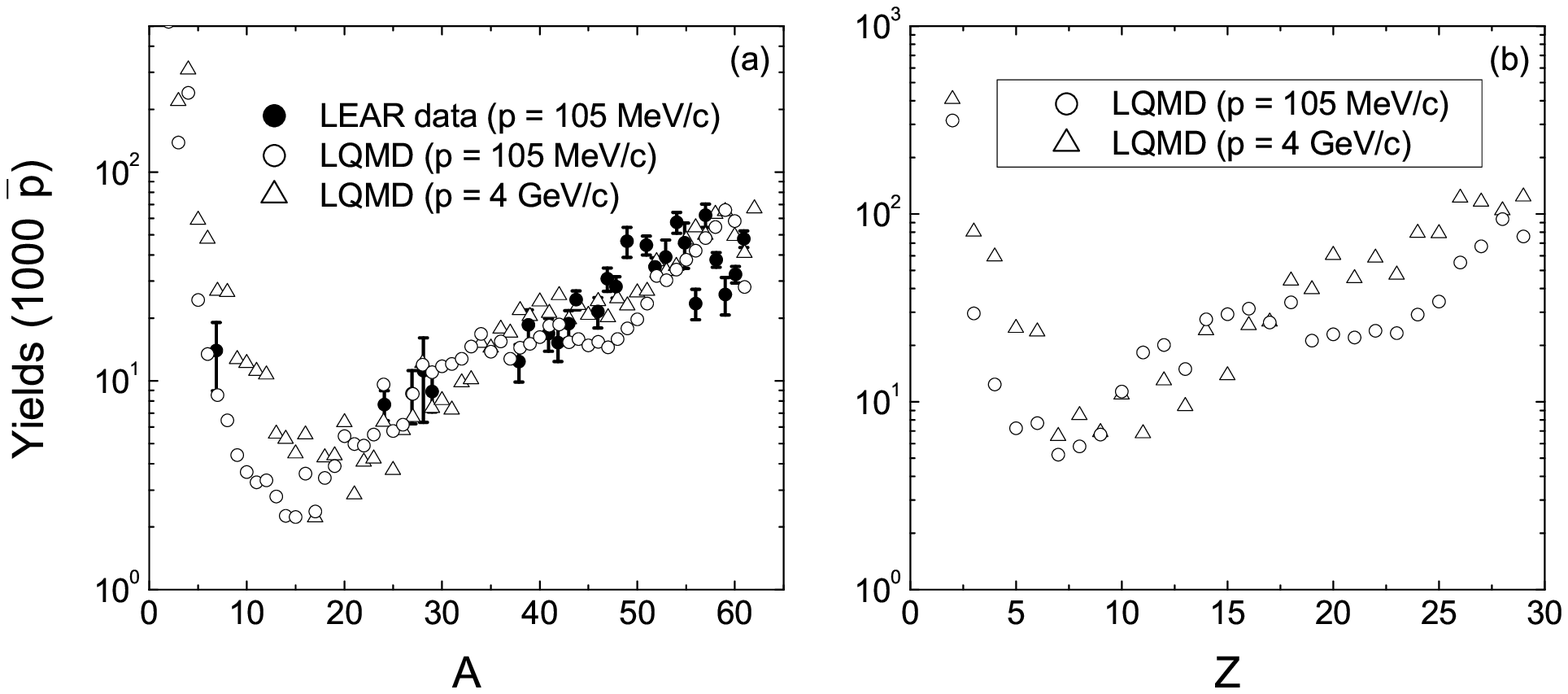}
\caption{ Mass and charge distributions of fragments produced in the $\overline{p}$ + $^{63}$Cu reaction at incident momenta of 105 MeV/c and 4 GeV/c, respectively. The mass yields are shown from LEAR facility at CERN \cite{Ja93}.}
\end{figure*}

\begin{figure*}
\includegraphics[width=16 cm]{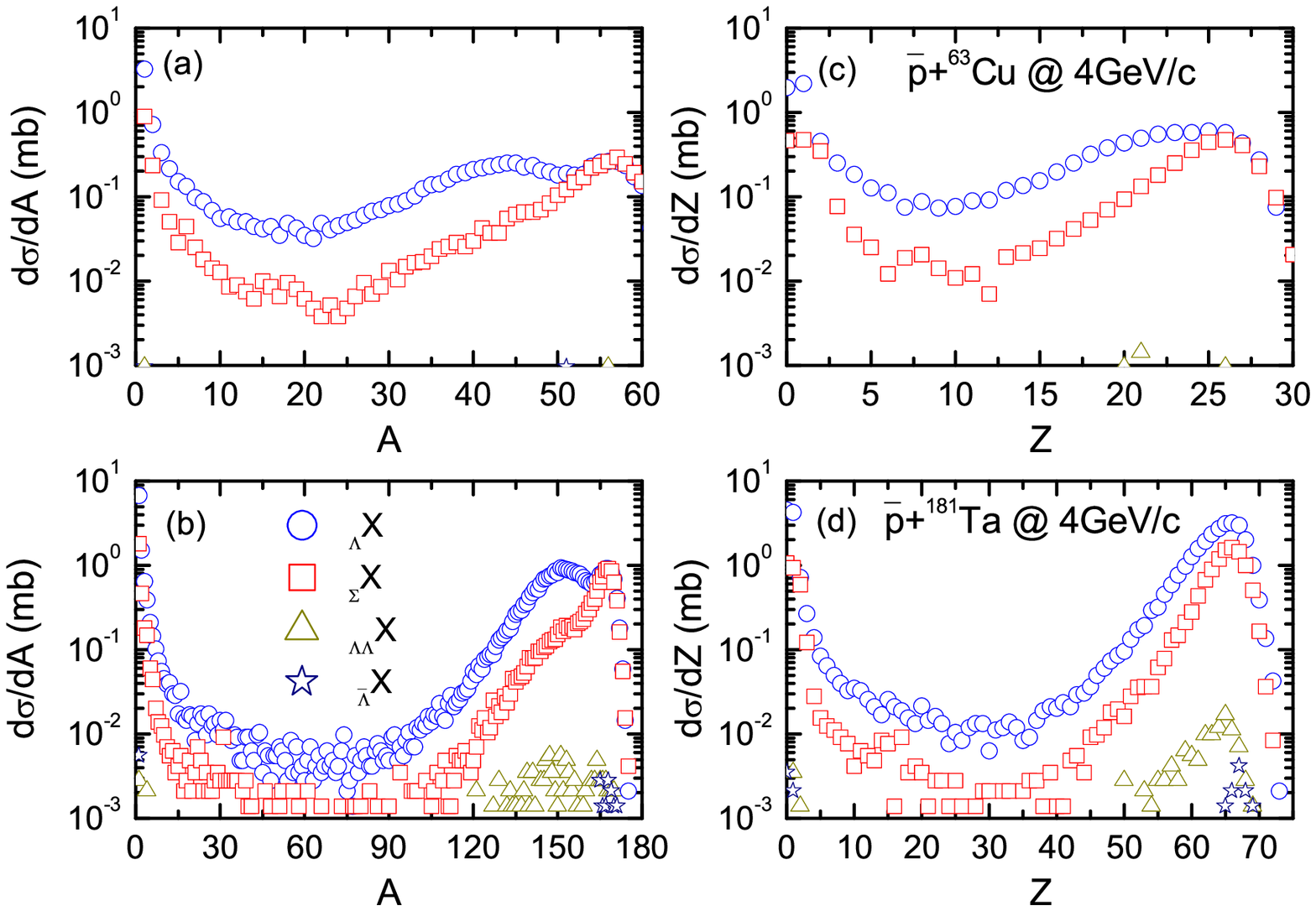}
\caption{(Color online) Hyperfragments with strangeness s=-1 ($_{\Lambda}$X), s=1 ($_{\overline{\Lambda}}$X) and s=-2 ($_{\Lambda\Lambda}$X) spectra as functions of atomic number (left panel) and charged number (right panel) in collisions of $\overline{p}$ on $^{63}$Cu (upper window) and $^{181}$Ta (down window) at incident momentum of 4 GeV/c.}
\end{figure*}

More information on the hyperfragment formation in antiproton induced reactions is pronounced from the mass and charged number distributions. Direct production of hypernuclei with strangeness s=-2 (double $\Lambda$-hypernucleus) and s=1 ($\overline{\Lambda}$-hypernucleus) in heavy-ion collisions or by proton induced reactions are difficulty because of very mall cross sections, in particular for the heavy-mass region. Properties of the hypernuclei would be significant in understanding the $\Lambda$-$\Lambda$ and $\overline{\Lambda}$-nucleon interactions, which are not well understood up to now. The ($K^{-}$, $K^{+}$) reactions are used for producing the double hypernucleus $^{6}_{\Lambda\Lambda}$He \cite{Ta01}. The antiproton-nucleus collisions would be a chance for producing the s=-2 and s=1 hypernuclei. As a test of the combined approach, shown in Fig. 4 is the mass and charge distributions in the fragmentation reaction of $\overline{p}$ + $^{63}$Cu at incident momenta of 105 MeV/c and 4 GeV/c, respectively. The available data from LEAR facility at CERN in the $\overline{p}$ + $^{\texttt{nat}}$Cu reaction\cite{Ja93} could be roughly reproduced. The products are averaged with 1000 antiprotons similar to the experimental condition. Non-equilibrium process in the collisions contributes the light fragment emission. Shown in Fig. 5 is the mass and charged number distributions of hyperfragments with strangeness s=-1 ($_{\Lambda}$X), s=1 ($_{\overline{\Lambda}}$X) and s=-2 ($_{\Lambda\Lambda}$X) in collisions of $\overline{p}$ + $^{63}$Cu and $\overline{p}$ + $^{181}$Ta at the same of incident momentum of 4 GeV/c. The $\Lambda$-hyperfragments spread the whole isotope range with the larger yields. The maximal cross sections for s=-1 and s=-2 hyperfragments are at the levels of 1 $mb$ and 0.01 $mb$, respectively. The lower production yields of $\overline{\Lambda}$-hyperfragments at the level of 1 $\mu b$ are found.

In summary, the formation mechanism of nucleonic fragments and hyperfragments in antiprotons induced reactions has been investigated within the LQMD transport model. The de-excitation of fragments is described with the help of the GEMINI approach. The fragmentation reactions induced by low-energy antiprotons can be nicely described with the combined approach. The production of hyperons is mainly attributed from the direct contribution of $N\overline{N}$ collisions, mesons induced and strangeness exchange reactions. Hyperfragments are formed within the narrower rapidities in comparison with nucleonic fragments and hyperons. Heavy hyperfragments close to the target-mass region have larger cross sections. The hypernuclei with strangeness s=-2 (double $\Lambda$-hypernucleus) and s=1 ($\overline{\Lambda}$-hypernucleus) would be feasible with the antiproton beams at PANDA (Antiproton Annihilation at Darmstadt, Germany) in the near future experiments.

\textbf{Acknowledgements}

This work was supported by the Major State Basic Research Development Program in China (2015CB856903), the National Natural Science Foundation of China Projects (Nos 11175218 and U1332207) and the Youth Innovation Promotion Association of Chinese Academy of Sciences.

\end{document}